\documentclass[aps,superscriptaddress,preprint]{revtex4}%
\usepackage{amsfonts}
\usepackage{amsmath}
\usepackage{amssymb}
\usepackage{graphicx}%
\usepackage{titlesec}
\titleformat*{\section}{\flushleft \bf \large}
\titleformat*{\subsection}{\flushleft \bf}
\titleformat*{\subsubsection}{\flushleft}
\begin{document}

\title{
Quantum oscillations in adsorption energetics of atomic oxygen on
Pb(111) ultrathin films: A density-functional theory study}

\author{Ziyu Hu}
\affiliation{College of Science, Beijing University of Chemical
Technology, Beijing 100029, People's Republic of China}
\affiliation{LCP, Institute of Applied Physics and Computational
Mathematics, P.O. Box 8009, Beijing 100088, People's Republic of
China}
\author{Yu Yang}
\affiliation{LCP, Institute of Applied Physics and Computational
Mathematics, P.O. Box 8009, Beijing 100088, People's Republic of
China}
\author{Bo Sun}
\affiliation{LCP, Institute of Applied Physics and Computational
Mathematics, P.O. Box 8009, Beijing 100088, People's Republic of
China}
\author{Xiaohong Shao}
\thanks{To whom correspondence should be
addressed. E-mail: shaoxh@mail.buct.edu.cn (X.S.);
zhang\_ping@iapcm.ac.cn (P.Z.)}
\affiliation{College of Science,
Beijing University of Chemical Technology, Beijing 100029, People's
Republic of China}
\author{Wenchuan Wang}
\affiliation{Laboratory of Molecular and Materials Simulation, Key
Laboratory for Nanomaterials of Ministry of Education, Beijing
University of Chemical Technology, Beijing 100029, People's Republic
of China}
\author{Ping Zhang}
\thanks{To whom correspondence should be
addressed. E-mail: shaoxh@mail.buct.edu.cn (X.S.);
zhang\_ping@iapcm.ac.cn (P.Z.)} \affiliation{LCP, Institute of
Applied Physics and Computational Mathematics, P.O. Box 8009,
Beijing 100088, People's Republic of China}

\begin{abstract}
Using first-principles calculations, we have systematically studied
the quantum size effects of ultrathin Pb(111) films on the
adsorption energies and diffusion energy barriers of oxygen atoms.
For the on-surface adsorption of oxygen atoms at different
coverages, all the adsorption energies are found to show bilayer
oscillation behaviors. It is also found that the work function of
Pb(111) films still keeps the bilayer-oscillation behavior after the
adsorption of oxygen atoms, with the values being enlarged by
2.10$\sim$2.62 eV. For the diffusion and penetration of the adsorbed
oxygen atoms, it is found that the most energetically favored paths
are the same on different Pb(111) films. And because of the
modulation of quantum size effects, the corresponding energy
barriers are all oscillating with a bilayer period on different
Pb(111) films. Our studies indicate that the quantum size effect in
ultrathin metal films can modulate a lot of processes during surface
oxidation.

\end{abstract}

\maketitle

\section{INTRODUCTION}

Because of the well-defined simple geometry, thin metal films that
are nanometers thick provides an ideal system to study the quantum
behaviors of electrons in reduced dimensions. Many researches have
been carried out to investigate the corresponding quantum size
effect (QSE) [i.e., the dependence on the film thickness] on
electronic properties of metal films. According to the
``particle-in-a-box'' model, the itinerant electrons confined in
metal films are quantized into discrete energy levels along the $z$
direction perpendicular to the film surface, giving rise to the
so-called quantum well states (QWS)
\cite{1Smith1996,2Paggel1999,3Thurmer2002,4Meyero1994,5A.Ka1993,6K.F1976,7T.C.2000,8Wei2002},
while the energy dispersion in the $xy$ direction remains parabolic,
forming a series of energy subbands. As the thickness of the metal
film increases, these subbands sequentially cross the Fermi level,
and make the physical properties of metal film oscillate
periodically. Among all the metal species, ultrathin Pb(111) films
have attracted vast attentions. By employing the epitaxy depositing
methods, atomically flat Pb(111) films can be perfectly fabricated
on both metal \cite{Thurmer2001} and semiconductor
\cite{34M.H.2004,Lin05} substrates. Moreover, extensive knowledge of
the electronic properties and crystallite shape have been attained
through systematical investigations. Promimently, it has been found
that the electron Fermi wavelength of Pb is nearly four times the
lattice spacing along the [111] direction, so most physical
properties of Pb(111) films such as the thermal stability, surface
energy and work function, all oscillate with the thickness in a
quasi-bilayer period \cite{8Wei2002,34M.H.2004,35P.C2005}. The
quantum effect of ultrathin Pb(111) films can also result in a
spectacular oscillatory superconductivity transition temperature, by
modulating the electron density of states near the Fermi level and
the electron-phonon coupling \cite{Guo2004}. Based on these previous
results, it is easily understood that besides these intrinsic
physical properties, QWS of ultrathin Pb(111) films will also affect
the surface reactions such as oxidation.

In fact, a lot of researches have been carried out recently to study
QSE on surface reactivities. Most of these studies are for
applications in catalysis, corrosion, and gas sensing. For example,
the lifetime of negative N$_{2}$ ions physisorbed on Ag(111) films
\cite{38P.J1999} and the desorption temperature of CO on Cu(100)
films \cite{39A.G2004} are both found to be modulated by the film
thickness. Apart from other reactivities, the oxidation of metal
surfaces needs special concerns since the  formed metal oxide layers
have dominated many crucial properties of metals in industrial
applications \cite{Henrich1994,Madix1994}. Previous experimental
studies have already revealed that the initial oxidation rate of
Mg(0001) \cite{Aballe2004} and Pb(111) films \cite{10Ma2007} depends
on the film thickness in an oscillatory way. So it is of high
interests to theoretically investigate the specific mechanisms of
the oxidation processes and how QSE influences the initial oxidation
rate. For the Mg(0001) system, unfortunately, present theoretical
studies have failed to explain its surface oxidation process. The
most notable is the long-term enigma of low initial sticking
probability of thermal O$_{2}$ molecules at Mg(0001), which has been
measured by many independent experiments
\cite{Driver1999,Aballe2004} but cannot be reproduced by adiabatic
state-of-the-art density functional theory (DFT) calculations
\cite{Hellman2005}. The central problem is that the adiabatic DFT
calculations were unable to find any sizeable barriers on the
adiabatic potential energy surface (PES), whose presence, however,
is essential for explanation of the experimental finding. This has
led to speculations that nonadiabatic effects may play an important
role in the oxygen dissociation process on the Mg(0001) surface. For
the Pb(111) system, fortunately however, the oxygen dissociation
process can be reasonably understood based on the adiabatic DFT
calculations. In fact, our previous studies have already revealed
that there are precursor adsorption states for molecular O$_{2}$,
and the molecular adsorption energy of O$_{2}$ shows a bilayer
oscillation QSE with the thickness of Pb(111) films
\cite{17Yangyu2008}. Moreover, we have also studied the adsorption
properties for atomic O both at surface and subsurface sites
systematically \cite{13Bosun2008}.

Based on these results, we further study the QSE on the adsorption
energies and diffusion energy barriers of oxygen atoms on Pb(111)
films in this work. The rest of the paper is arranged as follows. In
Sec. II we give details of the first-principles total energy
calculations, which is followed in Sec. III by our results of QSE on
the adsorption energy for on-surface and subsurface adsorption of
oxygen atoms. In Sec. IV, the QSE on the diffusion and penetration
of oxygen atoms on Pb(111) films is discussed. And finally, we give
our conclusions in Sec. V.

\section{COMPUTATIONAL METHOD}

The DFT total energy calculations are carried out using the Vienna
\textit{ab initio} simulation package \cite{VASP}. The generalized
gradient approximation (GGA) of Perdew \textit{et al}. \cite{PW91}
and the projector-augmented-wave (PAW) potentials \cite{PAW} are
employed to describe the exchange-correlation energy and the
electron-ion interaction, respectively. The so-called ``repeated
slab'' geometries \cite{Slab} are employed to model the clean
Pb(111) films ranging from 3 to 25 monolayers (ML). A vacuum layer
of 20 \AA~ is adopted to separate the Pb(111) slabs in two adjacent
cells, which is found to be sufficiently convergent from our test
calculations. The oxygen atoms are placed on both sides of the slabs
in a symmetric way. During our calculations, the positions of all
lead layers as well as the O atoms are allowed to relax until the
forces on them are less than 0.02 eV/\AA. The plane-wave energy
cutoff is set to 400 eV. A Fermi broadening \cite{FermiBroaden} of
0.1 eV is chosen to smear the occupation of the bands around ${\rm
E}_{F}$ by a finite-$T$ Fermi function and extrapolating to
$T\mathtt{=}0$ K. Integration over the Brillouin zone is done using
the Monkhorst-Pack scheme \cite{Pack} with $21\times21\times1$ grid
points at $\Theta$=$1.0$ coverage and $13\times13\times1$ grid
points at $\Theta$=$0.25$ coverage. The calculated lattice constant
of bulk Pb is 5.03 \AA, in good agreement with the experimental
value of 4.95 \AA~ \cite{Wyckoff}. Furthermore, the nudged elastic
band (NEB) method \cite{Mills} is used to find the minimum energy
path and the transition state for O atoms to penetrate or diffuse on
ultrathin Pb(111) films.

One central quantity tailored for the present study is the average
binding energy of the adsorbed oxygen atom defined as
\begin{equation}
E_{b}(\Theta)=-\frac{1}{N_{\text{O}}}[E_{\text{O/Pb(111)}}-
E_{\text{Pb(111)} }-N_{\text{O}}E_{\text{O}}],\tag{1}
\end{equation}
where $N_{\text{O}}$ is the total number of O adatoms present in the
supercell at the considered coverage $\Theta$ (we define $\Theta$ as
the ratio of the number of adsorbed atoms to the number of atoms in
an ideal substrate layer). $E_{\text{O/Pb(111)}}$,
$E_{\text{Pb(111)}}$, and $E_{\text{O}}$\ are the total energies of
the slabs containing oxygen, of the corresponding clean Pb(111)
slab, and of a free O atom respectively. According to this
definition, $E_{b}$ is also the adsorption energy $E_{\text{ad}}$
per O atom, i.e., the energy that a free O atom gains upon its
adsorption. Thus a positive value of $E_{b}$ indicates that the
adsorption is exothermic (stable) with respect to a free O atom and
a negative value indicates endothermic (unstable) reaction.

On the other hand, since in most cases, the oxygen chemisorption
process inevitably involves the dissociation of O$_{2}$ molecules,
thus the adsorption energy per oxygen atom can alternatively be
referenced to the energy which the O atom has in the O$_{2}$
molecule by subtracting half the dissociation energy $D$ of the
O$_{2}$ molecule,
\begin{equation}
E_{\text{ad(}1/2\text{O}_{2}\text{)}}=E_{b}-D/2.\tag{2}
\end{equation}
With this choice of adsorption energy, then a positive value
indicates that the dissociative adsorption of O$_{2}$\ is an
exothermic process, while a negative value indicates that it is
endothermic and that it is energetically more favorable for oxygen
to be in the gas phase as O$_{2}$. The binding energy of
spin-polarized O$_{2}$\ is calculated to be $D$=5.78\ eV per atom
and the O-O bond length is about 1.235 \AA.\ These results are
typical for well-converged DFT-GGA calculations. Compared to the
experimental \cite{Huber} values of 5.12\ eV\ and 1.21 \AA~ for O
binding energy and bonding length, the usual DFT-GGA result always
introduces an overestimation, which reflects the theoretical
deficiency for describing the local orbitals of the oxygen. We will
consider this overbinding of O$_{2}$\ when drawing any conclusion
that may be affected by its explicit value.

\section{Quantum size effect on the adsorption energy of atomic oxygen}

First, we do geometry optimizations for clean ultrathin Pb(111)
films with different thickness. This is necessary both for
verification of numerical reliability and for comparison with
adsorbed cases. The results for various physical quantities display
well-defined QSE, which is fully consistent with previous
first-principles report \cite{8Wei2002}, thus suggesting that the
present calculation is reliable and accurate.

Formally, there are four high symmetry sites for surface adsorption,
i.e., the top, bridge (bri), hcp, and fcc sites, which are
schematically shown in Fig. 1(a). In our previous study
\cite{13Bosun2008}, we have found that the top site is notably less
favorable than the fcc and hcp hollow sites. When the O atom is
placed on the bridge site, it always moves to the fcc hollow site
after relaxation. Actually, Fig. 6 below will show that the bridge
site is a saddle point in the O diffusion path from hcp to fcc site.
Combining with our another calculated result that after dissociation
of O$_{2}$ along the most stable channel, the two O atoms will
occupy the surface fcc and hcp hollow sites, respectively, we derive
that these two surface sites are most relevant and accessible for
experimental observation. Thus, our present numerical calculation of
QSE on surface adsorption is focused on the surface fcc and hcp
hollow sites.

Our calculated adsorption energies for the surface adsorption at the
fcc hollow site with the coverage of $\Theta$=$1.0$ ML are shown in
Fig. 2. Because of the similarities with the results at surface fcc
hollow site, the calculational results for the adsorption energies
at surface hcp hollow site will not be presented and discussed. One
can see from Fig. 2 that the adsorption energy mostly oscillates
with the film thickness in a bilayer period. However, this bilayer
oscillation is interrupted by a crossover at the film of 15 MLs,
which is noted as a beat in the bilayer oscillation behaviors and
comes from the slight difference between the electron Fermi
wavelength of Pb and four times of the lattice spacing along the
[111] direction \cite{Zhang2005}. In our calculations, we see that
the bilayer oscillation persists for films as thick as 21 MLs.

Between adsorbate and substrate, many physical properties such as
binding (adsorption) energy, amount of charge transfer, and work
function of the adsorption system change significantly with
coverage. So studying the change of these quantities at different
coverage has become one hot point. In this paper, our focus is to
explore the QSE modulating the materials' basic physical properties.
But we are still not very clear whether the oscillating behaviors of
materials' basic physical properties are changed at different oxygen
coverage. So, to get more detailed information about this problem,
we also calculate the surface adsorption energy of atomic oxygen at
the coverage of $\Theta=0.25$ ML, as a comparison with the results
at the coverage of $\Theta=1$ ML.

The calculated adsorption energy for the on-surface adsorption at
the 0.25 ML coverage is shown in Fig. 3. One can see that the
adsorption energy also oscillates with the film thickness in a
bilayer period. The difference from the result for the 1 ML coverage
lies in that the adsorption energy for the 0.25 ML coverage has a
crossover at the film of 10 MLs, and the oscillation orientation of
the adsorption energy is reversed for Pb(111) films of 3 to 11 MLs.
For Pb(111) films beyond 11 MLs, the bilayer-oscillation orientation
becomes the same, which is very interesting.

Another central concern for the adsorption of atomic species is the
change of the electronic structures of metal surfaces. So we here
further calculate the influence on the work function of Pb(111)
films from the adsorption of oxygen atoms. As shown in Fig. 4, the
work function of clean Pb(111) films oscillates with a bilayer
period, and encounters two crossovers at 4 and 13 MLs. This
phenomenon is identified as a 9-ML beating envelope on which the
bilayer oscillation is superimposed, and has been observed for a lot
of electronic properties of Pb(111) films. For example, the beating
on the bilayer oscillation of the surface energy happens at 8 and 17
MLs \cite{8Wei2002}, and that of the density of states at the Fermi
energy happens at 6 and 15 MLs \cite{10Ma2007}. After the 1 ML
on-surface adsorption of oxygen atoms, we can see from Fig. 4 that
the work function is enlarged by $2.10\sim2.62$ eV for all Pb(111)
films. It means that the surface electrons will encounter a higher
energy barrier to penetrate into the vacuum layer. Although the
values are enlarged, we see that the work function still remains the
bilayer oscillation behavior after the adsorption of oxygen atoms,
which is almost the same as the oscillation behavior in the work
function of clean Pb(111) films. Therefore, we deduce that the
change of the work function caused by the adsorption of oxygen atoms
is hardly influenced by the QSE of different Pb(111) films.

In parallel with the surface adsorption, we also calculate the QSE
on the adsorption energy for the subsurface adsorption of atomic
oxygen. As shown in Figs. 1(b)-(d), there are three different
high-symmetry sites for oxygen occupation in subsurface region. The
octahedral site (henceforth octa) lies just underneath the
on-surface fcc site, and one tetrahedral site (tetra-I) lies below
the on-surface hcp site. A second tetrahedral site (tetra-II) is
located directly below a first-layer metal atom. From the calculated
adsorption energies, we reveal that the most preferred adsorption
site for the oxygen atom is the tetra-II site, in a wide coverage
range \cite{13Bosun2008}. In fact, the adsorption state at the
tetra-II site is also found to be the most stable one in both
surface and subsurface adsorption states \cite{13Bosun2008}.
Therefore, to study the QSE on the adsorption energy of the tetra-II
adsorption state is very meaningful. Similar to the surface
adsorption, the subsurface adsorption energy for atomic oxygen is
also larger at a higher coverage \cite{13Bosun2008}. So we here
focus on the QSE for the 1 ML subsurface adsorption of oxygen atoms.

The result of the adsorption energy for subsurface oxygen is shown
in Fig. 5 as a function of the thickness of Pb(111) films, from
which one can clearly see a bilayer oscillation fashion. The
oscillation is interrupted respectively at the 7 and 16 MLs, proving
the existence of the 9-ML beating envelope. Although the bilayer
oscillation and the beating envelope are normal characters for a lot
of electronic properties of Pb(111) films, our calculational results
are in fact a little fortuitous, because the unconfined $p_x$ and
$p_y$ states of Pb also take part in the electronic interactions
with O atoms \cite{13Bosun2008}. And we are not surprised that the
beating behavior for the subsurface adsorption of oxygen atoms does
not accord well with that observed in surface oxidation
\cite{10Ma2007}. During the subsurface adsorption of O atoms, the
work function of Pb(111) films is found to be lowered down by about
$0.65\sim1.04$ eV.

\section{Quantum size effect on the diffusion and penetration of atomic oxygen}

Diffusion of oxygen atoms after dissociation is also an elementary
process during the whole surface oxidation process. Understanding
the diffusion process is also very important for many important
catalytic reactions such as oxidation of CO and hydrocarbons. So we
here present first-principles calculations to study the QSE on the
diffusion energy barriers of oxygen atoms on different Pb(111)
films. In our previous work, we have reported that the the diffusion
energy barrier is always larger for higher coverages adsorption of
oxygen atoms \cite{13Bosun2008}. And so we here mainly focus on the
QSE on the diffusion of surface oxygen atoms with coverage of 1 ML.
As shown in Fig. 6(a), the energy barrier for an adsorbed oxygen
atom to diffuse from the hcp (fcc) to the fcc (hcp) hollow site is
denoted as $\Delta E_1$ ($\Delta E_2$). The calculated energy
barriers for the diffusion of an adsorbed oxygen atom on different
Pb(111) films are shown in Fig. 6(b). One can see that $\Delta E_1$
is always smaller than $\Delta E_2$, indicating that the adsorption
state at the surface fcc hollow site is more stable than the one at
hcp hollow site for all Pb(111) films. The bilayer oscillation
behavior can be clearly seen from Fig. 6(b) for the both two kinds
of diffusion energy barriers ($\Delta E_1$ and $\Delta E_2$). The
beating envelope can also be seen for the 11 ML Pb(111) film.
Similar results of being modulated by the QSE of Pb(111) films are
also observed for the diffusion of subsurface oxygen atoms.
Therefore, we derive that QSE can modulate the diffusion behavior of
atomic species on ultrathin metal films.

To investigate the QSE on surface oxidation of Pb(111) films,
especially in more detail on the formation of surface oxide layers,
we also study the QSE on the penetrations of oxygen atoms after the
investigations of the QSE on diffusion energy barriers. It has
already been pointed out that the direct penetration for an oxygen
atom from the on-surface adsorption site into the tetra-II
subsurface site (the site right below a surface Pb atom) without
bypassing other subsurface sites is very energetically unfavorable
\cite{13Bosun2008}. So we here only consider the penetration of an
oxygen atom from the on-surface hcp (fcc) hollow site to the
neighboring subsurface tetra-I (octa) site at the coverage of
$\Theta=0.25$.

As shown in Figs. 7(a), the energy barrier for an oxygen atom to
penetrate into the subsurface tetra-I site is smaller than the
energy barrier for on-surface diffusions. So the adsorbed O atoms
are very possible to penetrate into subsurface sites. With the
energy barriers for an oxygen atom to penetrate into the subsurface
tetra-I and octa sites respectively denoted as $\Delta E_3$ and
$\Delta E_4$, one can see from Figs. 7(b) that $\Delta E_3$ is
always smaller than $\Delta E_4$ on different Pb(111) films,
indicating that the penetration from the on-surface hcp hollow site
is always easier. The energy difference between $\Delta E_3$ and
$\Delta E_4$ ranges between 0.22 and 0.27 eV for all studied Pb(111)
films. As functions of the thicknesses of Pb(111) films, we can see
from Figs. 7(b) that the energy barriers for the two penetration
paths both show bilayer oscillation behaviors. Therefore, in
addition to the modulation on the diffusion energy barriers, the QSE
also modulates the penetrations of adsorbed atomic species on metal
films.

\section{Conclusion}

In summary, we have systematically studied the adsorption and
diffusion energetics of oxygen atoms on different Pb(111) thin
films. The surface adsorption energies for atomic oxygen at
different coverages have been all found to be modulated by the QSE
of Pb(111) thin films. The bilayer oscillation behavior and the 9 ML
beating pattern have been identified. After surface adsorption of
atomic oxygen, it has been shown that the work function of adsorbed
Pb(111) thin films is enlarged, with the same bilayer-oscillation
mode as that of the clean Pb(111) thin films. The adsorption energy
for subsurface adsorption of oxygen atoms has also been found to be
modulate by the QSE and show a bilayer oscillation behavior.
Furthermore, we have studied the QSE of Pb(111) thin films on the
diffusion and penetrations energetics of adsorbed oxygen atoms,
which shows that the most energetically favored diffusion and
penetration paths are the same on different Pb(111) films. The
corresponding energy barriers have been all found to be modulated by
the QSE and display the prominent bilayer oscillations. Therefore,
we conclude from our present systematic results that the QSE in
ultrathin metal films can modulate a lot of important processes
during their surface oxidation.

\begin{acknowledgments}
This work was supported by the NSFC under grants No. 10604010 and
No. 60776063.
\end{acknowledgments}

\clearpage

\noindent\textbf{List of captions} \\

\noindent\textbf{Fig.1}~~~ (color online) (a) Four on-surface
adsorption sites including top, hcp, bri(bridge), and fcc sites.
(b), (c) and (d) The tetra-I, tetra-II, and octa subsurface
adsorption sites for oxygen atoms. The surface and second layer Pb
atoms are shown in black and grey balls, and the oxygen atoms are
shown in smaller red balls. \\

\noindent\textbf{Fig.2}~~~ (Color online) Calculated adsorption
energy for on-surface O atoms at the fcc hollow site at the 1 ML
coverage, as a function of thickness of the Pb(111) films. \\

\noindent\textbf{Fig.3}~~~ (Color online) The adsorption energy for
the 0.25 ML on-surface adsorption of oxygen atoms as a function of
thickness of Pb(111) films, in comparison with the adsorption energy
for the 1 ML adsorption. \\

\noindent\textbf{Fig.4}~~~ (Color online) Work function for clean
and O-adsorbed Pb(111) films as a function of the film thicknesses. \\

\noindent\textbf{Fig.5}~~~ (Color online) The adsorption energy for
subsurface O atoms at the tetra-II site of Pb(111) films as a
function the film thickness. \\

\noindent\textbf{Fig.6}~~~ (Color online) (a) Total energy for an
oxygen atom to diffuse on the 13 ML Pb film. (b) Energy barriers for
an adsorbed oxygen atom to diffuse on different Pb(111) films as a
function of the film thickness. \\

\noindent\textbf{Fig.7}~~~ (Color online)  (a) Total energy for an
adsorbed oxygen atom on the 13 ML Pb(111) film to penetrate from the
on-surface hcp hollow site to the subsurface tetra-I site and from
the on-surface fcc hollow site to the subsurface octa site. (b)
Calculated energy barriers for an adsorbed oxygen atom to penetrate
into the subsurface tetra-I and octa sites, as functions of the film
thickness. \\

\clearpage

\begin{figure}
\includegraphics[width=1.0\textwidth]{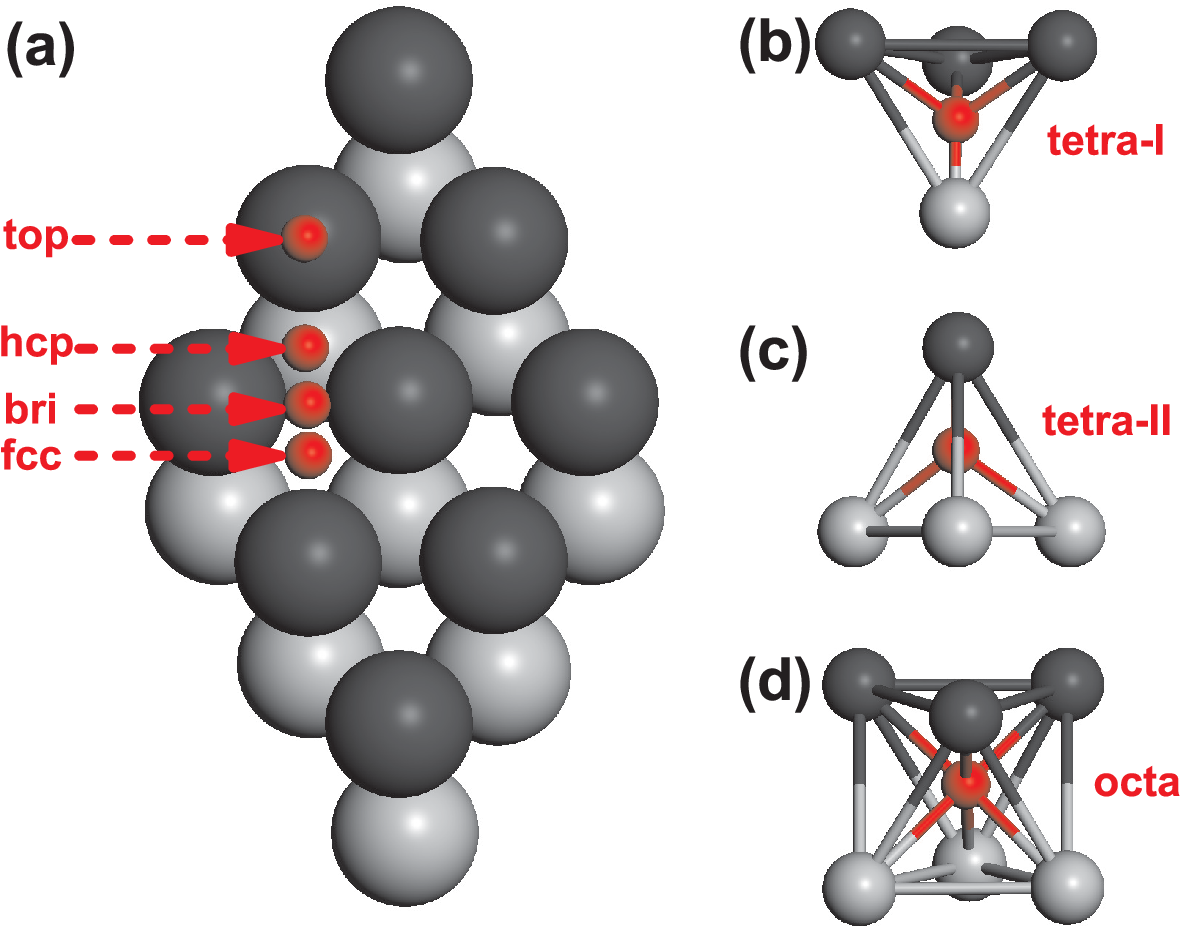}
\caption{\label{fig:fig1}}
\end{figure}
\clearpage
\begin{figure}
\includegraphics[width=1.0\textwidth]{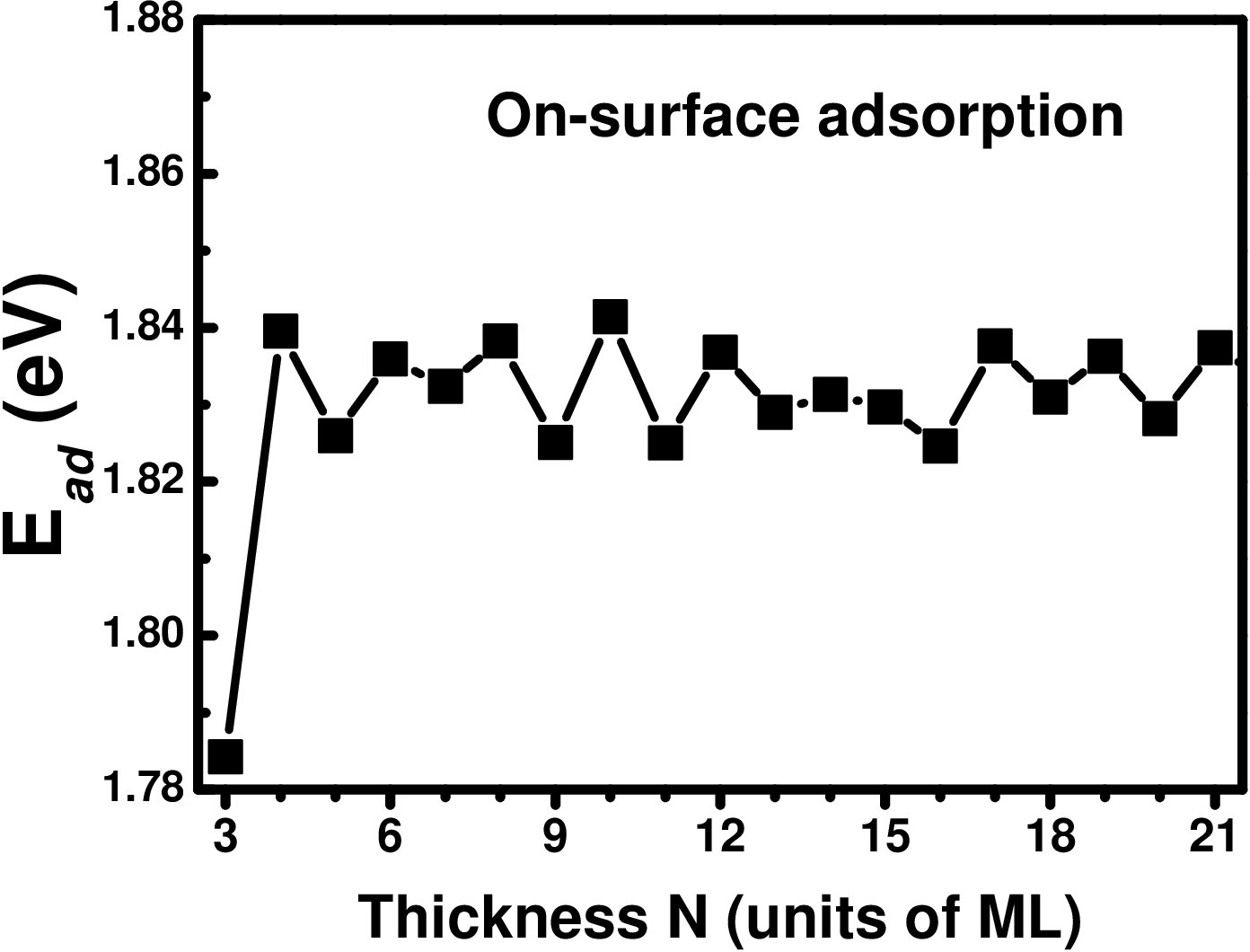}
\caption{\label{fig:fig2}}
\end{figure}
\clearpage
\begin{figure}
\includegraphics[width=1.0\textwidth]{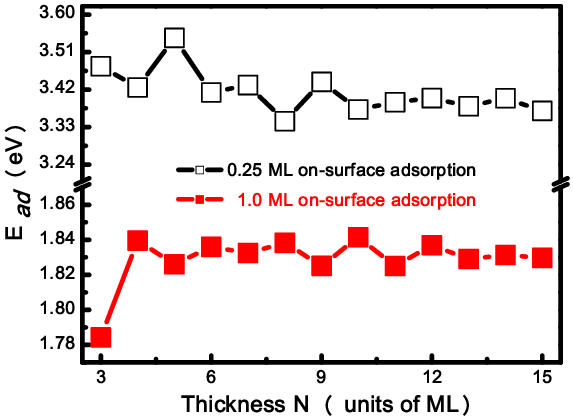}
\caption{\label{fig:fig3}}
\end{figure}
\clearpage
\begin{figure}
\includegraphics[width=1.0\textwidth]{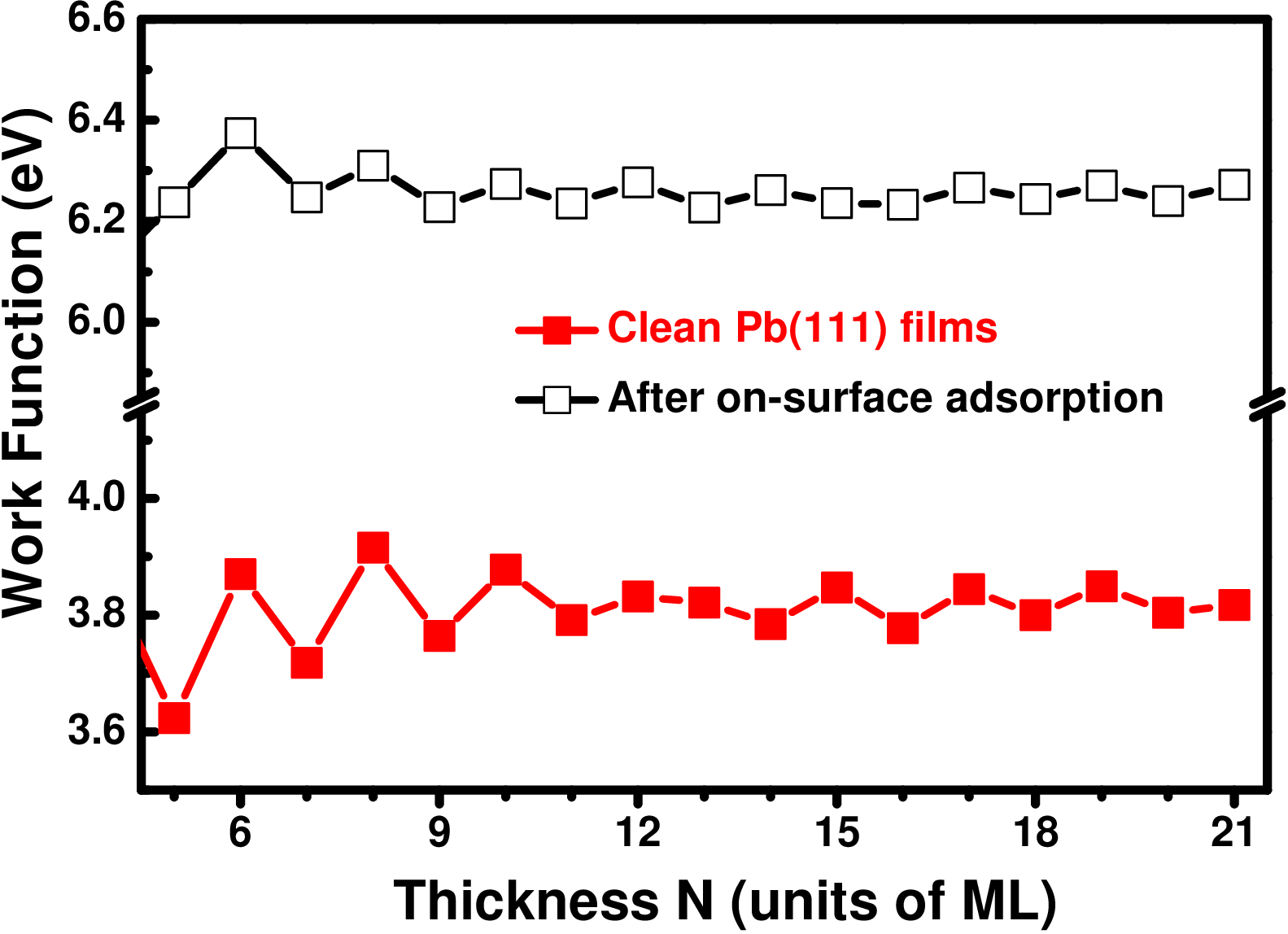}
\caption{\label{fig:fig4}}
\end{figure}
\clearpage
\begin{figure}
\includegraphics[width=1.0\textwidth]{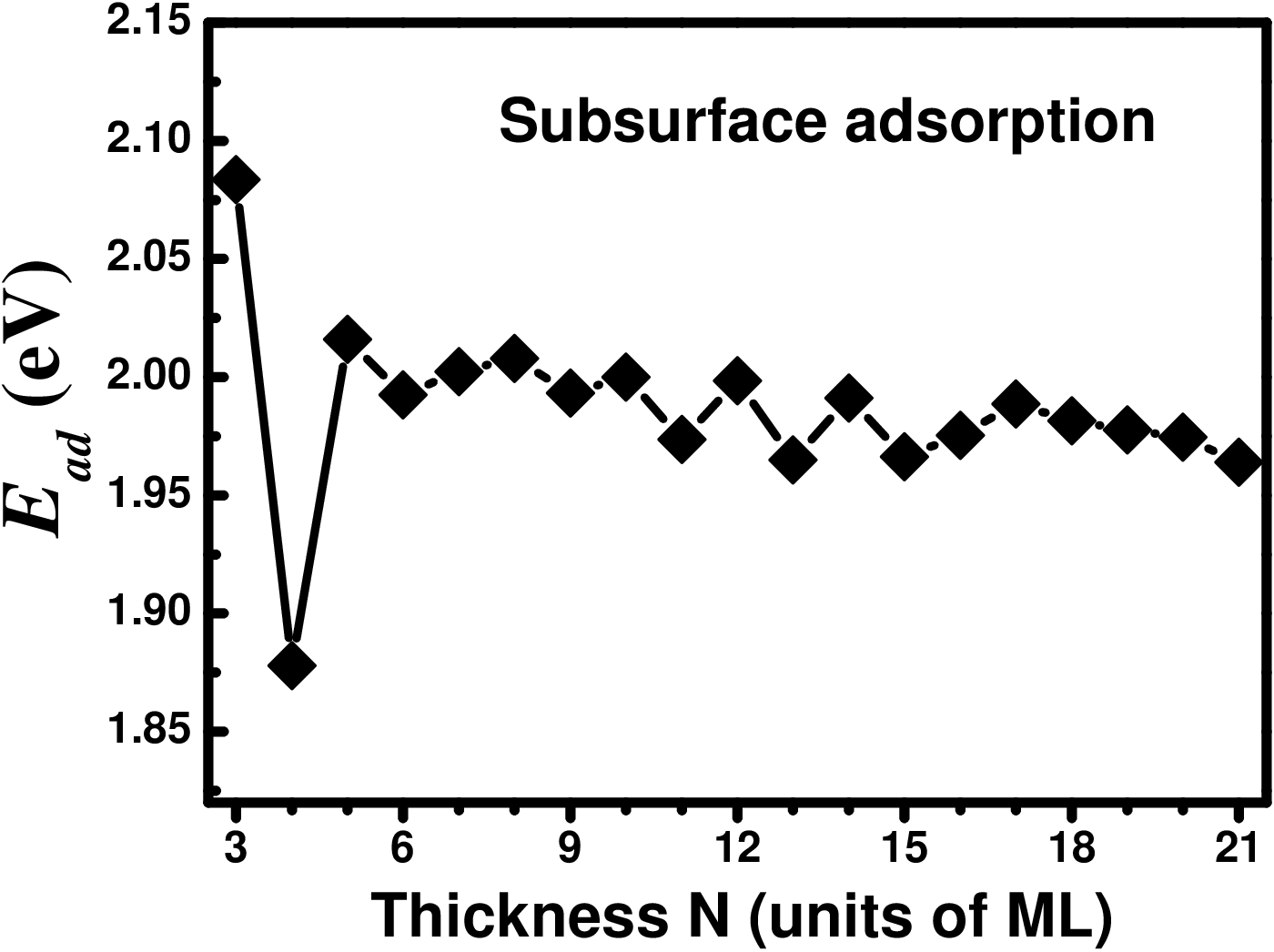}
\caption{\label{fig:fig5}}
\end{figure}
\clearpage
\begin{figure}
\includegraphics[width=1.0\textwidth]{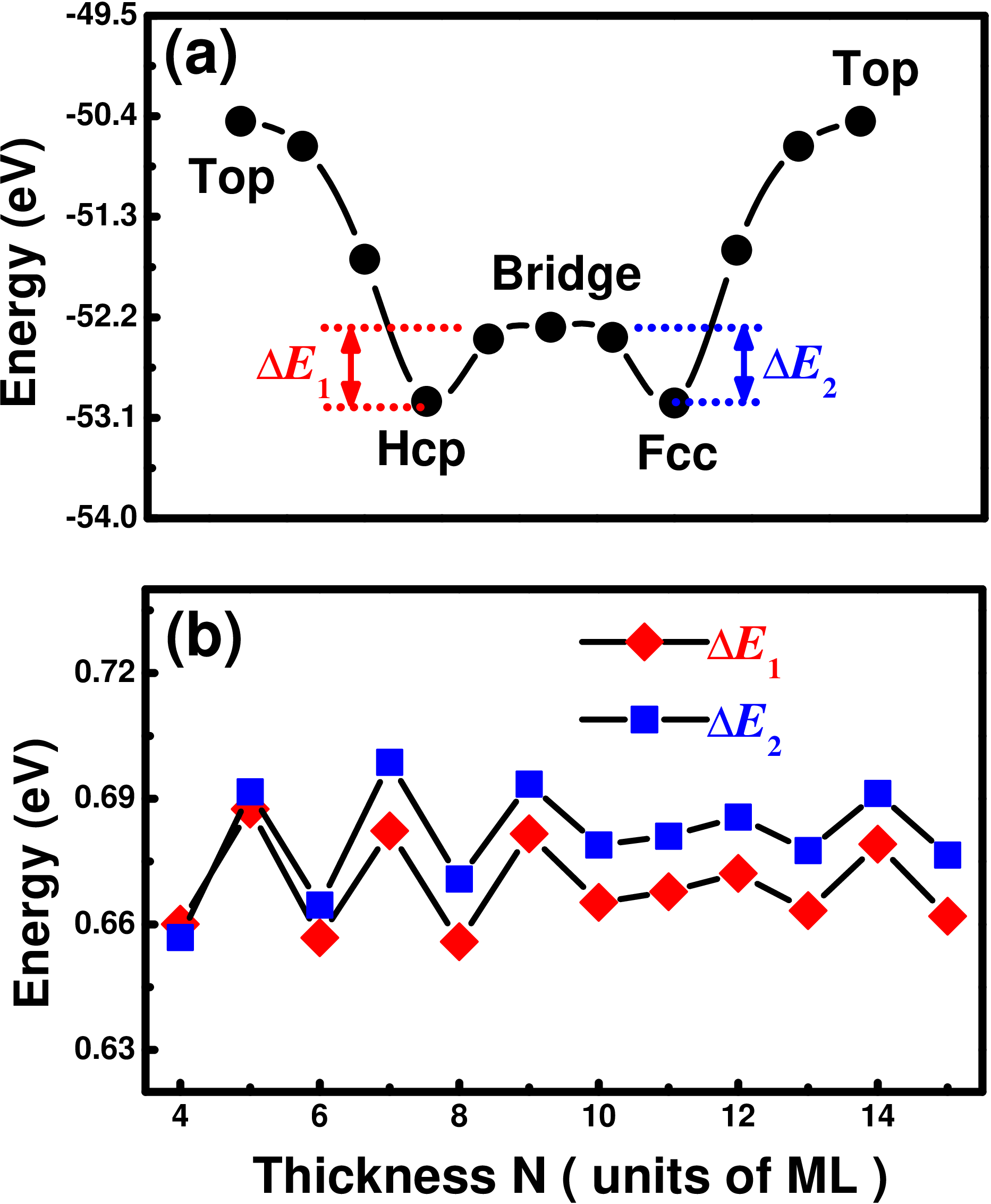}
\caption{\label{fig:fig5}}
\end{figure}
\clearpage
\begin{figure}
\includegraphics[width=1.0\textwidth]{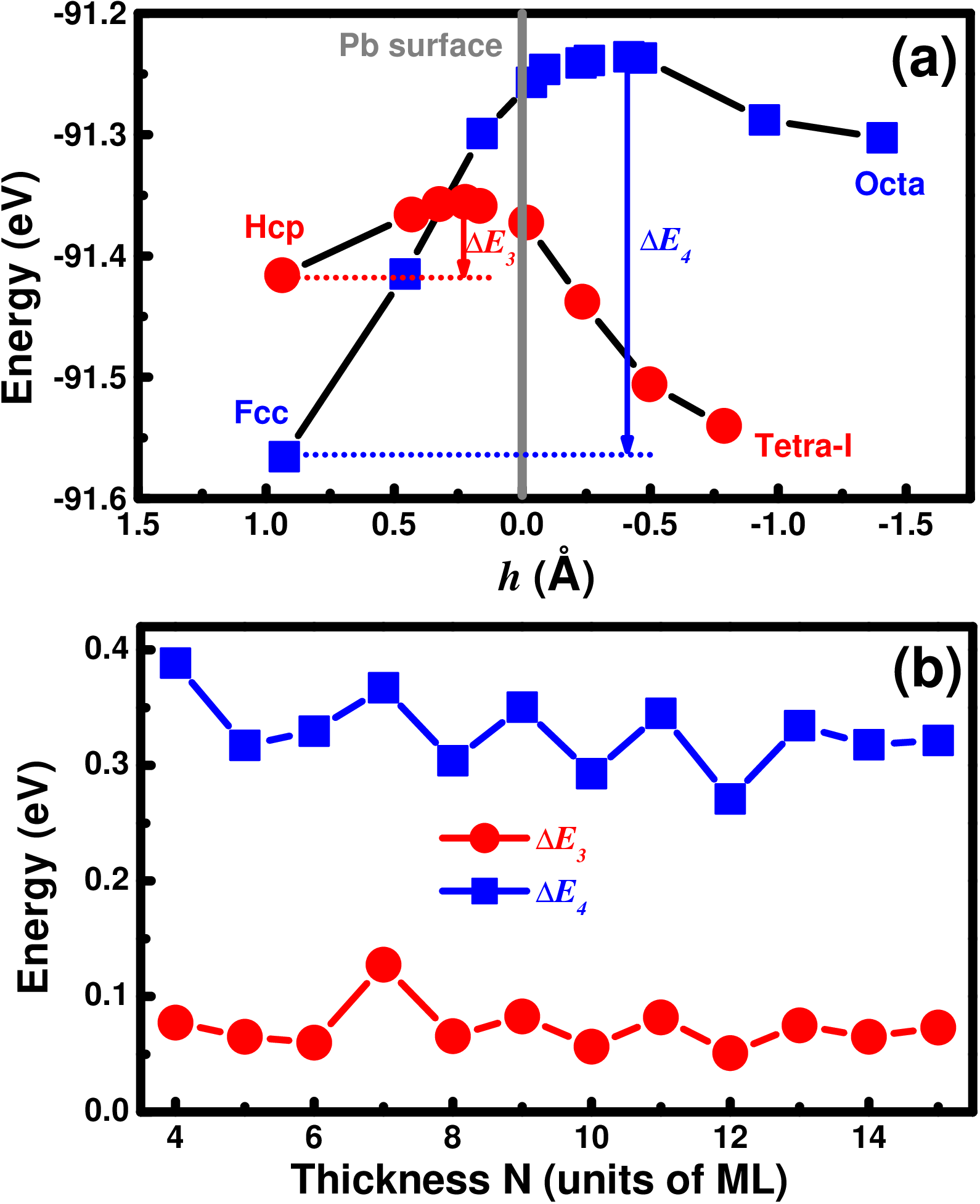}
\caption{\label{fig:fig5}}
\end{figure}
\end{document}